# Avoiding the "Great Filter": A Projected Timeframe for Human Expansion Off-World

Jonathan H. Jiang[1], Philip E. Rosen[2], Kristen A. Fahy[1]

1. Jet Propulsion Laboratory, California Institute of Technology, Pasadena, CA, USA
2. Energy Industry Engineer (Retired), Vancouver, WA, USA

## Abstract

A foundational model has been developed based on trends built from empirical data of space exploration and computing power through the first six plus decades of the Space Age which projects earliest possible launch dates for human-crewed missions from cis-lunar space to selected Solar System and interstellar destinations. The model uses computational power, expressed as transistors per microprocessor, as a key broadly limiting factor for deep space missions' reach and complexity. The goal of this analysis is to provide a projected timeframe for humanity to become a multi-world species through off-world colonization, and in so doing all but guarantees the long-term survival of the human race from natural and human-caused calamities that could befall life on Earth. Be-ginning with the development and deployment of the first nuclear weapons near the end of World War II, humanity entered a 'Window of Peril' which will not be safely closed until robust off-world colonies become a reality. Our findings suggest the first human-crewed missions to land on Mars, selected Asteroid Belt objects, and selected moons of Jupiter and Saturn can occur before the end of the 21st century. Launches of human-crewed interstellar missions to exoplanet destinations within roughly 40 lightyears of the Solar System are seen as possible during the 23rd century and launch of intragalactic missions by the end of the 24th century. An aggressive and sustained space exploration program, which includes colonization, is thus seen as critical to the long-term survival of the human race.

## 1. Introduction

Under existential threats and pressures such as sharp changes in population and demographics straining precious resources, climate change, nuclear war, pandemic, and the dramatic increase in the prevalence of facilities capable of producing deadly genetically engineered pathogens, it is easily understandable how one could become convinced humanity is living through an ever-expanding debris field of our own troubled history since the end of World War II. Is it just in our nature that we are bound to hurdle towards self-annihilation once the means are at hand, falling victim to the cosmic "Great Filter" [1]? The term "Great Filter" is a conceptualized probability threshold that could pose a barrier to the evolution of intelligent life.

Consideration of that question, which has challenged researchers since before the dawn of the Space Age, manifests analytically as the "L" factor (lifespan of technical/communicative civilizations) in the famed Drake Equation [2] and more recently found by statistical modeling to be the Equation's most influential term [3]. Indeed, "Our transient existence has lasted for less than 10 one-billionths of cosmic history so far on a tiny rock we call Earth, surrounded by a vast lifeless space" [4]. Against this backdrop, however, we are reminded that, "We are more than just our genes" [5]. Human minds sufficiently complex to render our home planet unlivable, have also created the technology able to travel through space and investigate other worlds. Given the fact the Earth also faces natural threats such as asteroid impacts, supernovae radiation, and super-volcanic activity, "our chance for survival could improve if some people choose to move away from Earth. Currently, all our eggs are in one basket. Venturing into space offers the advantage of preserving our civilization from a single-planet disaster" [6]. In recent decades, as robotic missions throughout the Solar System progressed, much speculation has been given to the next steps including colonization [7].

Focusing on the assurance for humanity's survival that would be afforded by permanent, self-sustaining, and genetically diverse colonization off-Earth, and deter-mining when such a Herculean task can be expected to occur, is paramount. Estimates for how long a technological civilization might last if confined to just its home world varies widely, from the pessimistic limit of destruction occurring in the immediate future, to "a few centuries but not much longer" [6] to approaching 8,000 years [8]. Such a broad range is to be expected given the complexities inherent in individual and group behaviors of humans across Earth's population of nearly 8 billion. Estimates for off-world colonization's timeframe can be expected to vary as well, functionally dependent as they are on the rate of technological development, government and private industry priorities, and public support, among other factors. However, the more than 60 years since the start of the Space Age have brought rapid development of humanity's off-world capability, from Sputnik in 1957 to men on the moon in 1969 to robotic probes having visited the far reaches of the Solar System and Kuiper Belt. Hence, a key parameter within humanity's control for impacting the odds of long-term survival can be found in how many years it will take to establish at least one permanent, self-sustaining and genetically diverse off-world colony. This 'window of peril', which opened in 1945 for the technological civilization of Earth, will remain a threat until such colonies become a reality.

But how do we meet the challenge of estimating how long until robust off-world colonies are established? Answering, even in the very approximate, such a far-reaching query requires extensive modeling involving both quantitative and qualitative data from the start of the Space Age to the present, plus projections into the coming decades of the 21st century that encompass an array of unknowns such as findings from the Moon, Mars, Jovian and Saturnian satellites already underway in this decade. Rather than trying from the outset to build a complex multi-variable model, the first steps of an incremental approach is presented here whose intent is to lay the foundation for building the complex model by first keying on a single critical parameter. The choice for this critical parameter is computing power, and is driven by the fact of its own exponential technological development has run parallel to that of space exploration technology. Cutting edge space exploration technology historically relies on computational capability for handling multitudes of complex calculations which enables rapid progress. For the purposes of this analysis, computing power will be measured as the number of transistors that can be contained within a single microprocessor. The logic of this choice follows directly from the clear dependence of scientists, inventors, engineers, medical researchers and others on their ability to gather, process and interpret the vast quantities of information required to conceive, design, build, test, operate and maintain the new technologies required for increasingly complex and distant human missions into deep space. "[Shawn] Domagal-Goldman and [Giada] Arney, [NASA Goddard astrobiologists] envisage future exoplanet missions where AI technologies embedded on spacecraft are smart enough to make real-time decisions, saving the many hours necessary to communicate with scientist on Earth" [9].

At this point it should be noted that computing power, expressed as transistors per microprocessor, does have an analogy to the human brain. More specifically, the human brain contains ~1011 neurons and ~1014 synaptic connections among those neurons [10]. Further, "A transistor can be thought of as an analog to the synapse (connection) between two neurons" [11]. As well, with cutting edge technology such as using 'persistent photoconductivity' found in some perovskite semiconductors as a kind of 'optical memory', this "…phenomenon can also mimic synopses in the brain that are used to store memories" [12]. Hence, computing power as described by transistor count per micro-processor can be thought of as a cousin to the human brain at the basic structural level, with both organic and synthetic computational devices being used in concert to develop the new technologies of the Space Age.

Finally, it should be noted that any estimate of deep space mission timing merely projects a date of earliest launch from Earth (or, more generally, cis-lunar space, which means "not beyond the moon" in Latin), but does not include actual travel time to a given mission's final destination or the follow-on time interval to establish human colonies once the first crewed landing is made. On this note, while chemically-powered rockets can likely suffice for missions to Mars, Jovian space and, perhaps, as far as the Saturn system, their inherent exhaust velocity limitation of about 5 km/sec precludes their consideration for practical interstellar travel [13]. Travel to interstellar destinations would require highly advanced technology such as generational star ships and/or breakthrough propulsion systems – e.g., Bussard Ramjet [14] utilizing a CNO-catalyzed nuclear fusion cycle [13, 15]. In any case, greater computing power than available today will be needed to engineer, build and operate such designs.

## 2. Methodology

As computing power grows at an ever-accelerating rate, approximately doubling every two years during much of the second half of the 20th century [16] and continuing into the 21st century, so too has humanity's reach into outer space. The robotic component of the Space Age, which began with Sputnik in 1957, quickly advanced to the first lunar flybys in 1959 (Luna 1 and 2), flyby missions to Venus (Mariner 2, 1962), Mars (Mariner 4, 1964), a successful soft landing on the Moon (Luna 9, 1966), and launches of later successful flybys of the outer planets by Pioneer 10 in 1972, Pioneer 11 in 1973 and Voyager 1 and 2 in 1977. The model discussed below is based on the confluence of these two exponential trends: computing power (expressed as transistors per microprocessor) as a function of time and distance (measured in astronomical units, AUs) from launch point Earth to the first robotic deep space missions' farthest successfully achieved objective as a function of time.

## 2.1 Computational Power and Deep Space Missions Over Time

As computing power grows at an ever-accelerating rate, approximately doubling every two years during much of the second half of the 20th century [16] and continuing into the 21st century, so too has humanity's reach into outer space. The robotic component of the Space Age, which began with Sputnik in 1957, quickly advanced to the first lunar flybys in 1959 (Luna 1 and 2), flyby missions to Venus (Mariner 2, 1962), Mars (Mariner 4, 1964), a successful soft landing on the Moon (Luna 9, 1966), and launches of later successful flybys of the outer planets by Pioneer 10 in 1972, Pioneer 11 in 1973 and Voyager 1 and 2 in 1977. The model discussed below is based on the confluence of these two exponential trends: computing power (expressed as transistors per microprocessor) as a function of time and distance (measured in astronomical units, AUs) from launch point Earth to the first robotic deep space missions' farthest successfully achieved objective as a function of time.

2.1. Computational Power and Deep Space Missions Over Time

As will be seen, both trends can be easily plotted in a semi-log manner with the linear horizontal axis being the number of years since the start of the Space Age (Figure 1). Exponential trend curves are then best-fit to the plotted data, yielding:

$$C = 15.70226 \exp(0.34821\, T) \qquad (1)$$

where C = computing power, T = time in number of years since 1957, and the correlation coefficient $R^2$ = 0.994. See Figure 1 left-panel for the graphical representation of data corresponding to equation (1).

And: $$D = 0.04846 \exp(0.31272\, T) \qquad (2)$$

where D = distance from launch point Earth to farthest successfully achieved mission objective in AUs (minimal distance from Earth's orbit to objective's orbit), T = time in number of years since 1957, and the correlation coefficient $R^2$ = 0.994. See right-panel of Figure 1 for the graphical representation of data corresponding to equation (2). Please refer to Figure 3 at the end of this section for a simplified flow chart of the development of this and other relationships which comprise the core of our model.

With the commonly time dependent equations of these two data sets in hand, it is then a straight forward exercise to combine the two equations as to eliminate T, their time dimension, creating a new power relationship between computing power and distance from launch point Earth to farthest successfully achieved robotic mission objective:

$$C = 457.2437\, D^{1.11334} \qquad (3)$$

where the correlation coefficient $R^2$ = 0.996. It should be noted that the computing power vs. time data set contains 24 data points over the 61 years between 1957 and 2017 [17] while the deep space missions vs. time data set includes only 5 data points. This limitation, implicit in its reliance on the relatively small number of deep space missions' empirical data, cannot be avoided.

Equation 3 provides a relationship between how far distance-wise, robotic deep space missions have progressed relative to concurrent computing power. This relation between a fundamental element which underlies the technology and procedures necessary for venturing into deep space for extended periods of time, and ultimately establishing off-world colonies, is coupled to the physical reach of such efforts.

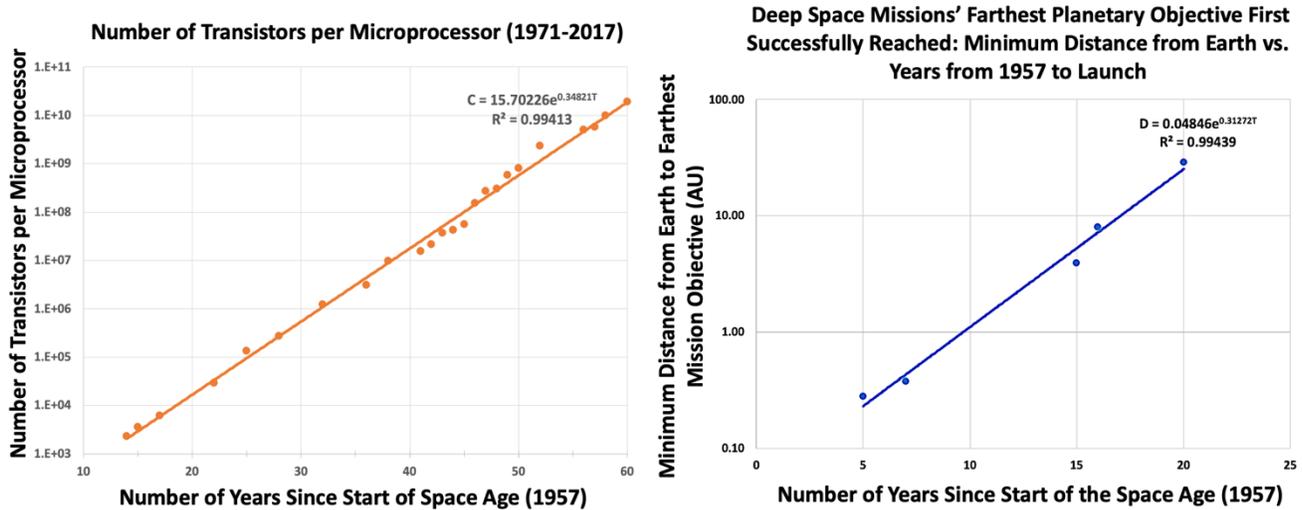

**Figure 1:** Left-panel; Computational power as expressed in number of transistors per microprocessor from 1971 to 2017, time expressed as number of years since start of the Space Age (1957), used in empirically derived equation (1) [17]; Right-panel: First of their kind robotic deep space missions' (flybys) planetary objective minimum distance from Earth orbit (in astronomical units) versus number of years from start of Space Age (1957) to missions' launch year, used in empirically derived equation (2).

While it may appear more straightforward to estimate the timing of additional and more extensive deep space missions by basing such projections solely on selected successful missions already flown – i.e., simply following the established trend further out in space and time, doing so would blunt the influence in any such estimate from the critical component of concurrent technological development. Indeed, when just the simple logarithmic extrapolation of first successful flyby missions actually flown is used as the basis – i.e., equation (2), solved for time as a function of distance, to estimate timing of far-reaching human-crewed missions, an ever-widening discrepancy arises with respect to increasing mission objective distance. As an example, for the nearest interstellar mission to Proxima Centauri, the aforementioned simplistic extrapolation predicts a first successful launch date of 2080 for human-crewed mission, 174 years sooner than predicted by our model which takes into account as well the projected time required to develop mission-necessary computing power. When one considers that Moore's Law is expected to run up against material and engineering limitations [18], while radical new forms of computing to carry out ever more complex calculations (e.g., quantum computing) are yet to be perfected, the means to accomplish the developmental dimension of humanity's most daunting endeavors cannot be ignored.

Since the first successful interplanetary robotic flyby missions in the first half of the 1960s, Mariner 2's visit to Venus and Mariner 4's close encounter with Mars, rapid advancement in computing have accompanied and supported ever more far reaching and complex missions. These include flybys of all the outer planets, Pluto and a Kuiper Belt Object, orbiters in the Jovian and Saturnian systems, asteroid regolith sampling and a series of semiautonomous rovers scouring the surface of Mars for signs of an ancient life-supporting environment. While human-crewed missions have yet to venture beyond the Moon, advances in control and automation systems, materials and engineering techniques, as well as in space medicine (largely thanks to the database amassed from two decades of long-term crew postings to the International Space Station) made in the half century since Apollo have paved the way for humans to feasibly set foot on Mars before the close of the 2030s.

## 2.2 From Robotic Missions to Human-Crewed Landings

Robotic missions have and will continue to pave the way for human crews to follow. However, the leap from relatively inexpensive (and ultimately expendable) machine voyagers to craft capable of keeping groups of highly complex living organisms alive in the most hostile of environments is an enormous leap in nearly every design and construction detail. As one aspect, humans evolved under the protection of Earth's strong magnetic field and dense atmosphere which work in combination to screen-out the vast majority of high energy particles from interstellar space – commonly referred to as cosmic rays. Direct exposure to cosmic rays, even for relatively short periods of time, would pose grave health risks to an unprotected crew. Although shielding can be employed to meet this need, this would add substantially to the engineering requirements of crewed spacecraft along with additional mass necessitating more powerful engines, which in turn adds further to the mass of the vehicle presenting engineers with what amounts to a Newtonian conundrum. Nonetheless, the first 10 years of the Space Age brought spectacular example of bridging robotic to human landings, this beginning with Luna 1 and 2's flyby of the Moon in 1959, Luna 9's soft landing in 1966, and culminating in 1969 with the first crewed landing by Apollo 11. While this progression provides a critical data point for this analysis, at least one other data point would be needed to begin proposing a quantifiable time-distance relationship between robotic and human missions. It is at this point a degree of speculation is needed concerning humanity's next likely target for 'boots on the regolith' – Mars. Mariner 4 first flew by Mars in 1964 and was followed by the first successful soft landers, Viking 1 and 2, in 1976. While humans have yet to step foot on the Martian surface, a first crewed landing in the 2030s is envisioned [19]. For the purposes of this analysis, we will venture a somewhat conservative assumption that such a landing successfully launches in 2038. Note that while the optimal Hohmann transfer orbit launch windows for a Mars mission in the late 2030s will actually occur in 2037 and 2039, for the purposes of our calculations and average of 2038 will be used). When going from launch of a first actual or first projected flyby mission to a first robotic lander mission for a given destination, a fixed number of years is simply added to the actual or projected flyby launch date which corresponds to the actual delta realized for Mars, that of 11 years (Mariner 4's flyby in 1964 to Viking 1 and 2's launch in 1975). The implicit assumption is that the technological leaps required to go from robotic flybys to robotic landings are minor compared to going from robotic flyby to human landings. Further, for robotic missions the intermediate step between flybys and landings, that of an orbiter without a lander, is ignored per the leap from flyby to orbiter also being minor when compared to the challenges of placing a complex device safely on the surface of another world. Finally, while several of the Soviet era Venera probes to Venus did successfully place robotic landers on the surface, the severe conditions char-acteristic of Venus greatly limited their operating window and would almost certainly rule out Venus as a viable destination for human landings, much less colonization, in the foreseeable future. For these reasons only Mariner 2's Venus flyby in 1962 is included in the construction of our model.

With two sets of data points for time interval and distance when going from robotic flyby to human landing missions, noting again that the second set of data points, this involving Mars, are partially speculative, it is possible to propose a mathematical relationship. While any curve can be perfectly fitted to this small a data set, this analysis will confine itself to a logarithmic relationship between time (expressed as number of years between first successful robotic flyby and human landing) vs. distance from launch point Earth to mission objective (measured in AUs). The resulting equation is as follows:

$$T_{fh} = 12.824 \ln(D) + 86.532 \tag{4}$$

where $T_{fh}$ = number of years from launch of first successful robotic flyby mission to launch of first successful human landing. In the case of the Moon, D = 0.00256 AU while $T_{fh}$ is 10 years. For Mars, D = 0.376 AU (closest orbital approach, not necessarily the distance a crewed or non-crewed vehicle would take in the instance of a Hohmann transfer orbit) while $T_{fh}$ is projected to be 74 years.

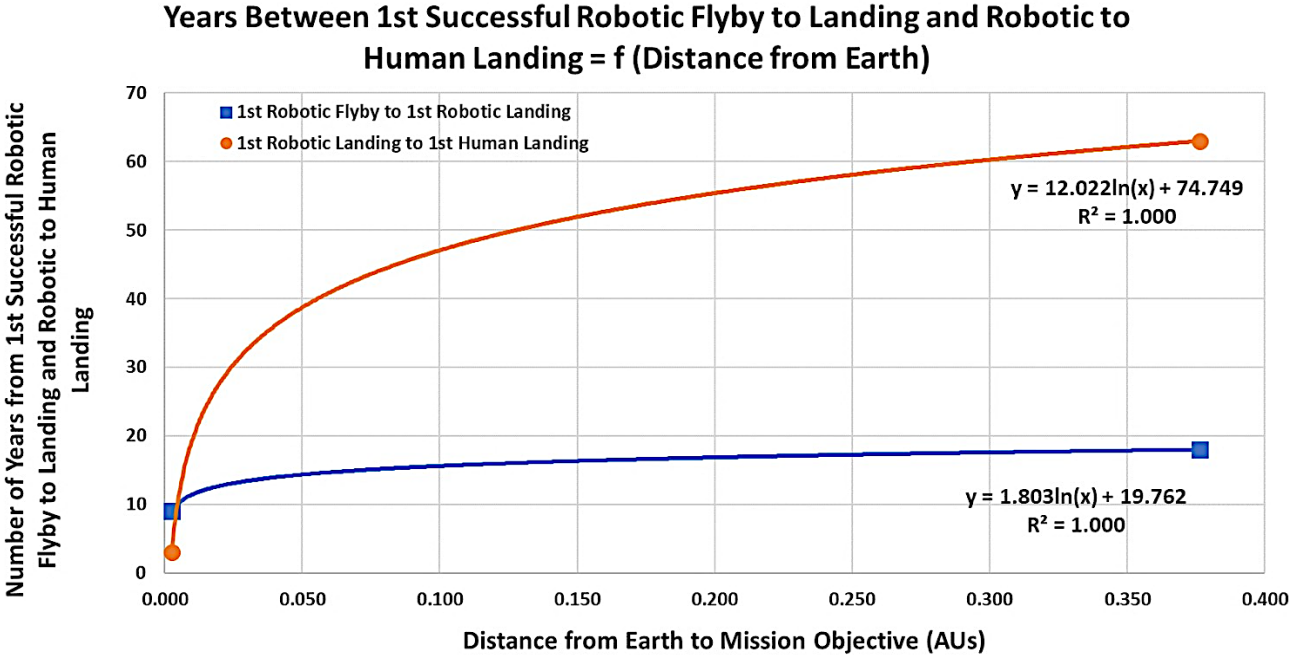

**Figure 2:** Number of years separating launch of first successful robotic flyby to robotic landing and first successful robotic landing to human landing, for Moon and Mars missions. Note: Launch of first successful human landing on Mars is assumed to occur in calendar year 2038.

The rationale for using a logarithmic relationship is founded in the nature of technological development itself, specifically the familiar S-curve of development vs. time. First, there is a relatively slow run-up over decades which can be regarded here as our going from the Wright Brothers' first powered flight of a heavier than air vehicle in 1903 to Sputnik in 1957 – i.e., the lower portion of the S-curve. Technological progress then experiences a sharp upward acceleration over a short period of time, this covering the first ~20 years of the Space Age, that took us from Sputnik to Voyager 1 and 2 – i.e., the steep middle portion of the S-curve. Finally, with much of the 'low hanging fruit' of technical barriers having been surmounted for relatively near-Earth mission objectives, the long and gently curving upper section of the S-curve is encountered. It is the upper half of the S-curve that is relevant to our analysis and best expressed mathematically in the logarithmic form of equation (4) and as illustrated below in Figure 2.

Finally, there is consideration for the time interval in going from the first crewed landing to the first self-sustaining, genetically viable (i.e., robust) colony. As no such colony has yet been established off-world, actual data for folding into this analysis is pending though the coming years with the prospect for the beginnings of said colonies (NASA, 2019). For the purposes of this analysis, it will be assumed that once the planned Artemis Base Camp is successfully established near the Moon's south polar region, launches for true colonization can begin within a relatively short time frame from the successful first crewed landings' launch to Mars and beyond.

## 2.3 Assembling the Model

Figure 3 outlines how the above empirically derived equations are combined to create the model for extended deep space exploration. It must again be emphasized that computing power, while a broadly affecting factor, is only one such factor in a highly complex interplay of dynamic influences on how far and how soon humanity will establish off-world colonies. As such, this model is best considered a starting point version of more rigorous versions to come (see 5. Future work discussion).

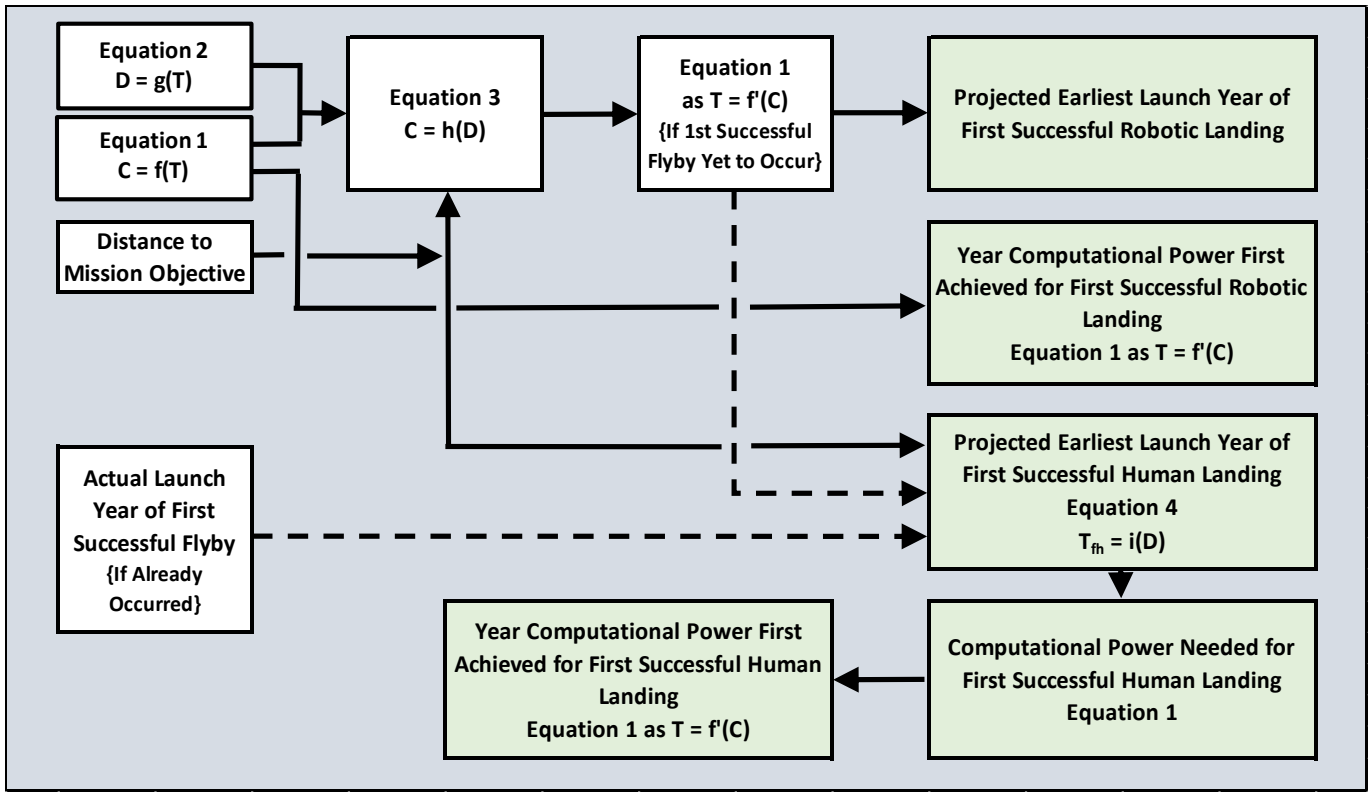

**Figure 3**: Empirical equations 1 and 2, in combination with distance to mission objective and flyby mission launch year data, yield projections for extended missions' launch years and computational power required. Note: "f", "f'", "g", "h" and "i" denote functional relationships.

## 3. Results and Discussion

The data and derived equations contained herein estimate the projected calendar year dates and associated computational power that will be required for the successful launch of initial robotic landing and human landing missions from Earth to a given set of deep space mission objectives.

### 3.1 Actual and projected timing for deep space missions and associated computational power

In Table 1, the Solar System mission destinations were chosen per their history of robotic visits, and in the case of the Moon, the human landings that have taken place. Note again the assumption for first human landing on Mars to launch from Earth in year 2038 (averaged from 2037 and 2039). The rationale for the interstellar mission destinations are as follows:

**Proxima Centauri** – The closest known star to the Solar System at 4.2 lightyears, this active red dwarf is strongly suspected of having a least two planets and at least one of those, Proxima Centauri b, in the star's habitable zone [22].

**Tau Ceti** – The closest known G-class (Sol-like) star not part of a tandem star system, it is 11.9 lightyears away and hosts at least four planets and at least one in the star's habitable zone (http://phl.upr.edu/press-releases/twonearbyhabitableworlds).

**Trappist System** – A cool red dwarf star 40.5 lightyears away that is thought to host at least seven terrestrial-sized planets and several among those potentially in the star's habitable zone [23].

**4 kpc from Center of the Milky Way** – An annular region located approximately 4,000 parsecs from the center of the Milky Way Galaxy (closest point: ~14,000 lightyears from the Solar System, in the direction of Sagittarius and the galactic center) has been suggested by statistical modeling as the region of the Milky Way most likely to have contained complex life, including technological civilizations, in the past [3].

**Table 1:** Actual and projected timing of first successful robotic and human missions for selected destinations within the Solar System and interstellar space.

| Tables for LookUp Data on Pre-Chosen Destinations and Projections for Earliest Possible Mission Launch Dates | | | | | | | | | |
|---|---|---|---|---|---|---|---|---|---|
| **Destination** | Moon | Mars | Asteroid Belt | Jovian System | Saturn System | Proxima Centauri | Tau Ceti | Trappist System | 4 kpc from CoMW |
| **Dist. from Earth (AU)** | 0.0026 | 0.3763 | 1.5587 | 3.9501 | 8.0412 | 265,486 | 752,526 | 2,562,570 | 882,424,035 |
| **1st Flyby** | 1959 | 1964 | 1972 | 1972 | 1973 | Not Yet Launched | Not Yet Launched | Not Yet Launched | Not Yet Launched |
| **1st Robotic Lander** | 1966 | 1975 | Not Yet Launched | Not Yet Launched | 1997 | Not Yet Launched | Not Yet Launched | Not Yet Launched | Not Yet Launched |
| **1st Human Landing** | 1969 | 2038 | Not Yet Launched | Not Yet Launched | Not Yet Launched | Not Yet Launched | Not Yet Launched | Not Yet Launched | Not Yet Launched |
| **Actual and Calculated Projections for Calendar Year of Launch from Earth of First Successful Mission Type and Destination and Corresponding Computational Power Requirement** | | | | | | | | | |
| **Launch Year of 1st Successful Flyby Mission** | 1959 | 1964 | 1972 | 1972 | 1973 | 2007 | 2010 | 2014 | 2032 |
| **Year Computational Power First Achieved** | 1948 | 1964 | 1968 | 1971 | 1973 | | | | |
| **Launch Year of 1st Successful Robotic Landing** | 1959 | 1975 | 1979 | 1982 | 1984 | 2018 | 2021 | 2025 | 2043 |
| **Year Computational Power First Achieved** | 1959 | 1975 | 1979 | 1982 | 1984 | 2018 | 2021 | 2025 | 2043 |
| **Launch Year of 1st Successful Human Landing** | 1969 | 2038 | 2064 | 2076 | 2086 | 2254 | 2270 | 2290 | 2383 |
| **Year Computational Power First Achieved** | 1969 | 2038 | 2064 | 2076 | 2086 | 2253 | 2269 | 2289 | 2381 |

Utilizing the data in Table 1, Figure 4 provides a graphical representation of when, in terms of the Common Era calendar year, the first successful launches from Earth have occurred and are estimated to occur for both robotic and human landings to interplanetary and interstellar mission destinations.

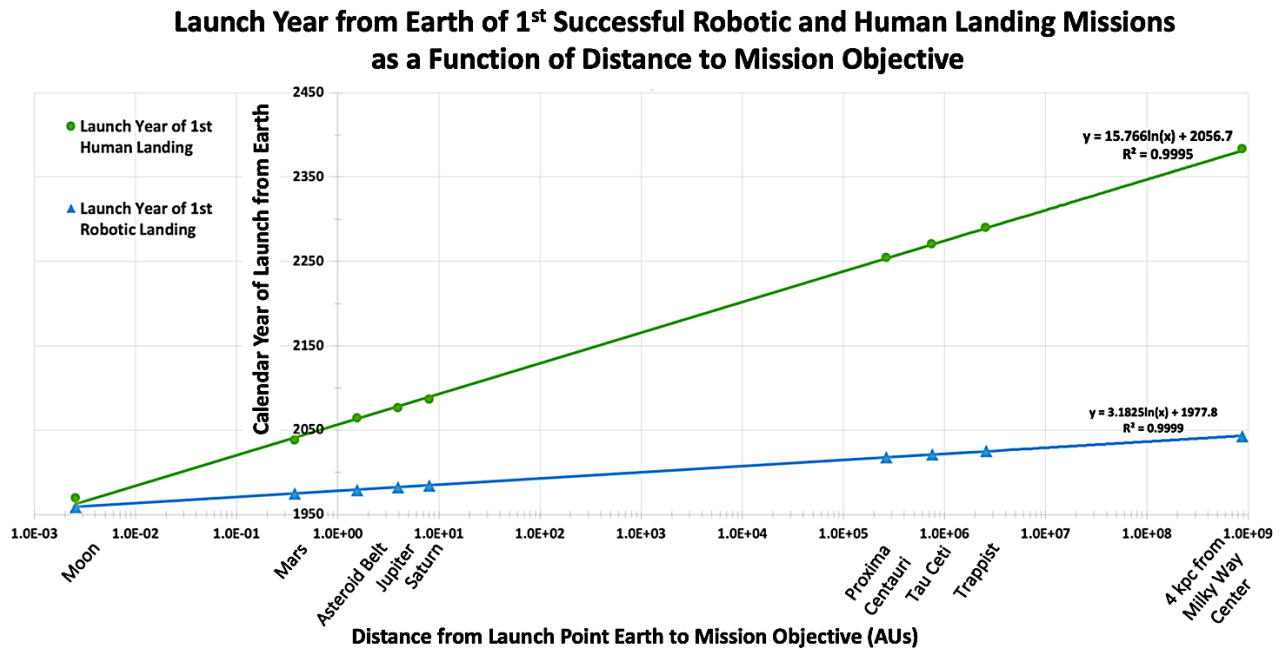

**Figure 4:** Graphical representation of data in Table 1 for launch date from Earth/cis-lunar space of robotic and human landing missions having destinations ranging from the Moon to approximately 14,000 lightyears from the Solar System.

### 3.2 Interpretation of selected missions' results

The above results are calculated for nine pre-chosen mission destinations, the first five of which are within the Solar System and representative of actual mission destinations flown starting in the late 1950s while the remaining four destinations are interstellar and run the range from our nearest stellar neighbor to approximately halfway to the center of the Milky Way Galaxy. Continued is a brief discussion for each destination:

**Moon:** The world closest to Earth and, so far, the only world humans traveled to and have landed upon. Having already been visited by humans, this places the Moon somewhat removed from the intent of this analysis given much of the most challenging aspects of its colonization have either been solved or are now coming into reach in the very near term with NASA's proposed 'Artemis Base Camp' by the late 2020s [21], as well as plans by other nations (e.g., China) to establish crewed facilities on the lunar surface in the coming decades. Given that the Moon is only a four-day trip from LEO by existing spacecraft technology, resupply from Earth would be far easier than even the next closest world for colonization, Mars, which currently takes a minimum of roughly six months to reach.

**Mars:** As stated in Section 2.2, an assumption has been made for a first successful human landing on Mars to occur in the late 2030s, this likely coming as an extension of the Artemis Program [21].

Hence, Mars is effectively treated as a fixed point in our analysis with speculation that at least the beginnings of self-sustained colonization can be expected well before the end of the 21st century. Supporting this assumption is analysis suggesting Mars to be the most eco-nomically viable location in the Solar System for colonization [24] and the focus of at least one well-funded private venture.

**Asteroid Belt:** Popular media references to human missions to the Asteroid Belt, which lay between the orbits of Mars and Jupiter, often involve mining operations. This is not without some scientific backing such as the potential for substantial metals recovery from 16 Psyche which is the target of a NASA robotic orbiter mission set to launch in August 2022. Our model predicts first human launch to an Asteroid Belt object could happen by the mid-2060s, with speculation that any human colonies to follow may eventually take hold on some of the larger bodies such as Ceres, 2 Pallas or 4 Vesta.

**Jovian System:** Our model predicts a first human-crewed launch to Jovian space may be possible by the mid-2070s. A more optimistic assessment was assembled by NASA in a 2003 study which detailed a human-crewed launch to the outermost Galilean moon, Callisto, by the mid-2040s [25]. While the four Galilean moons Io, Europa, Ganymede, and Callisto are the most obvious targets for robotic exploration, only the outermost moon, Callisto, is sufficiently removed from Jupiter's intense radiation fields to consider human exploration and possible colonization within the 21$^{st}$ century. Further robotic explorations of the Jovian system are planned such as the European Space Agency's Jupiter Icy Moons Explorer, set to launch in 2022, and NASA's Europa Clipper, scheduled to launch in 2026 - the findings from which may well further the case for humans to follow or suggest other, more viable, destinations.

**Saturn System:** Saturn and its many moons, while roughly twice as distant from Earth as the Jovian system, is projected to be reachable via a first human-crewed launch that would take place in the mid-2080s. Saturn's largest moon, Titan (possessing of a thick atmosphere of light hydrocarbons and the only interplanetary moon to have been visited thus far by a robotic lander), is an obvious target for long range plans owing largely to its vast hydrocarbon resources. However, Enceladus is at least as intriguing for human colonization due to the suspected subsurface ocean, as indicated by observations of recurrent $H_2O$ plumes erupting from its south polar region [26]. While the Jovian moon Europa is also strongly suspected of having a subsurface ocean, the aforementioned radiation concerns may make Enceladus a superior target for human colonization.

**Interstellar Destinations:** As discussed in Section 3.1, the first crewed launch to even the nearest interstellar destination, the Proxima system, is not envisioned until the mid-23rd century. Once, however, this great leap has been sent on its way, launches to star systems 'moderately' more distant such as Tau Ceti at nearly 12 lightyears away are projected to be possible within another two to three decades. A first human launch to still more distant solar systems such as Trappist at approximately 40 lightyears from the Solar System would entail another roughly two decades. Finally, the launch of a human-crewed voyage on the galactic scale, such as traversing the Milky Way's disk to reach halfway to its center, remains a goal much more distant in time and would not be expected possible before the late 24$^{th}$ century.

It must again be emphasized that all projections for human-crewed missions described above are first possible launch dates from Earth/cis-lunar space. Without revolutionary breakthroughs in propulsive engineering, human voyages – and any colonization that would follow – to interstellar destinations is highly unlikely to be undertaken. Fortunately, the Solar System presents a target rich interplanetary environment for astronauts to further explore and adventuring colonists to conquer starting here in the 21st century.

## 4. Conclusions

A simple model based on trends built from empirical data of space exploration and computing power through the first six plus decades of the Space Age has been developed which projects earliest possible launch dates of human-crewed missions to Solar System and interstellar destinations. The model algebraically folds in the expected broadly limiting factor of computational power, as quantitatively expressed in transistors per microprocessor, to the demonstrated extent of robotic and human missions to the Earth's Moon and the other planets and moons of the Solar System. While space exploration for purely scientific ends is a noble goal and certainly worthy in and of itself of the costs and sacrifices of such great endeavors, the goal of this analysis is to provide a timeframe for humanity to become a multi-world species through off-world colonization that would logically follow on the heels of earlier human landings. By creating at least one, and preferably multiple, self-sustaining and genetically viable off-world colonies, humanity can best assure its long-term survival from both natural and human-caused calamities that threaten life on Earth. The immediacy of such threats began with the development of the first nuclear weapons and has proliferated in multiple directions in the three quarters of a century since the end of World War II. In the early 21st century we find ourselves in the midst of a 'Window of Peril' which, having been opened in 1945, cannot be safely closed until said colonies are realized.

Per the findings of our model, launches from Earth of the first human-crewed missions to land on Mars, selected minor planet/planetoid bodies in the Asteroid Belt, at least one of the four Galilean moons of Jupiter, and at least one of the moons of Saturn should be possible before the end of the 21st century – if computational power is the only limiting factor and advances in microcircuitry continue apace. Human-crewed missions to interstellar destinations, while theoretically possible, will be dependent not only on development of onboard AI capability but also technologically advanced propulsive systems, the development of which is logically dependent as well on an ever-greater ability for computation. If undertaken and pursued to success, humanity will have not only all but assured its own survival but will have become a true spacefaring civilization [27]. Off-world colonization, like the Great Filter's self-annihilation, is within humanity's reach – which we choose as our future will constitute Earth's reply to Fermi's Paradox.

## 5. Future Work

The methodology for deriving the projected results of this analysis suggests at least two avenues for further expansion. Referring to Figure 5 (a generalization of Figure 3), below, they are:

1. Folding-in additional 'Associated Empirical Functions' (i.e., other affecting factors for equation (1) such as the trend of annual inflation-adjusted public and private sector funding for space exploration programs) to the 'Primary Empirical Function' of actual missions flown, expressed above as equation (2). This would yield a new version of the 'Modified Primary Empirical Function' for equation (3) through which the new 'Associated Empirical Function' is expressed in a revised 'Objective Function'. In addition to the funding parameter, other empirically derived associated functions may include but are not limited to international cooperation (building on the historic successes of early robotic missions to Venus and Mars, Apollo-Soyuz, and the International Space Station), public support, prior robotic survey findings, political and social priorities. Note that while some of these parameters may prove difficult to quantify in the direct sense, usable trends for the purpose of this type of analysis may still be extracted from data covering the six plus decades of the Space Age. From here, multiple 'Objective Functions' would be generated that would then need to be either algebraically combined to a 'Global Objective Function' or, more simply, their individual 'Projected Results' consolidated via weighted averaging scenario.
2. Adding an iterative feature to the calculations in which initial and intermediate model results are sequentially compared against the data set for the 'Primary Empirical Function'. Any discrepancies above a chosen tolerance would then generate successive correction factors that would be folded into the 'Objective Function' for each iteration until 'Projected Results' are brought within the chosen tolerance.

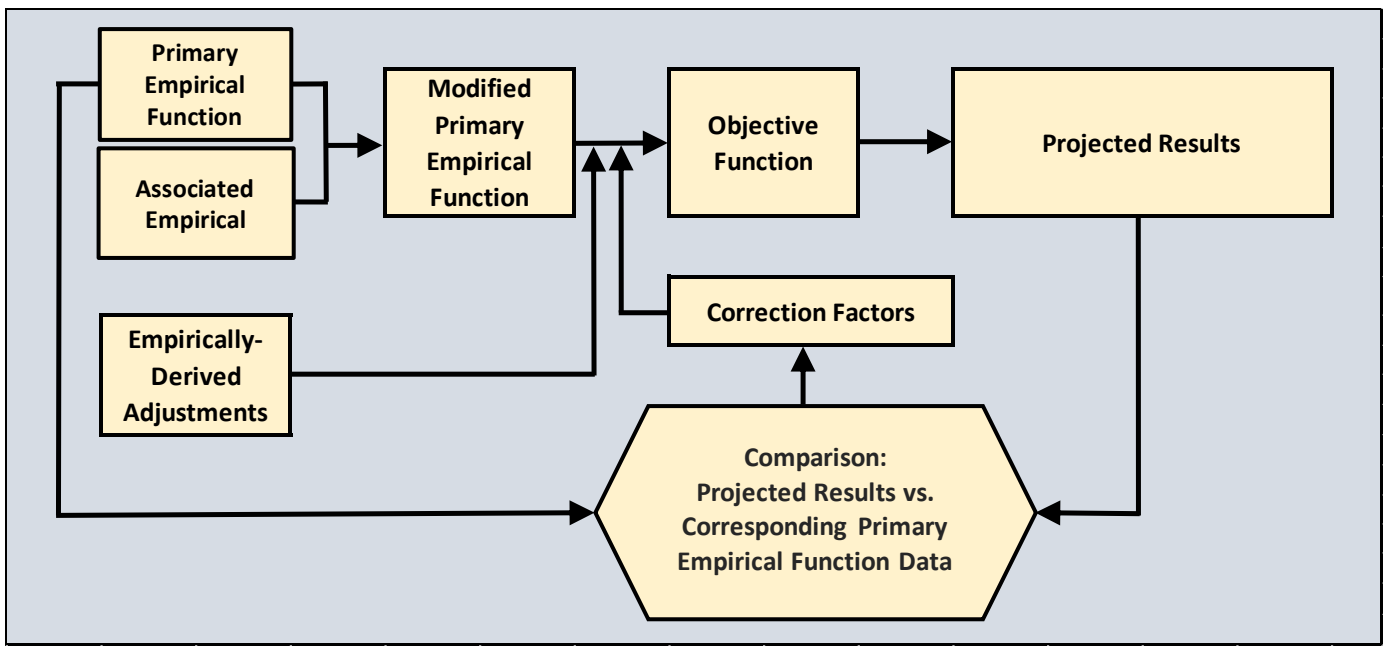

**Figure 5:** Methodology for combining an empirically derived primary function with an empirically derived associated function (future work item 1), plus optional recursive loop for successive corrections (future work item 2).

**Acknowledgments**

This research was supported by the Jet Propulsion Laboratory, California Institute of Technology, under the contract with NASA. We acknowledge the partial funding support from the NASA Exoplanet Research Program NNH18ZDA001N.

**Data Statement**

The data underlying this article are available in the article and in its online supplementary material. For additional questions regarding the data sharing, please contact the corresponding author at Jonathan.H.Jiang@jpl.nasa.gov.

**References**

[1] Hanson R. (1998) The Great Filter – Are We Almost Past It? Downloaded from: (https://web.archive.org/web/20100507074729/ http://hanson.gmu.edu/greatfilter.html), accessed May 13, 2021

[2] Drake F.D. (1965) The radio search for intelligent extraterrestrial life. In *Current aspects of exobiology* pp. 323-345.

[3] Cai X., Jiang J.H., Fahy K.A., Yung, Y.L. (2021) A Statistical Estimation of the Occurrence of Extraterrestrial Intelligence in the Milky Way Galaxy. *Galaxies* 9(1), 5

[4] Loeb A. (2020) A Sobering Astronomical Reminder from COVID-19, *Scientific American*, downloaded from: (https://www.scientificamerican.com/article/how-much-time-doeshumanity-have-left/), Accessed April 18, 2020.

[5] Hawking S. (2018) Brief Answers to the Big Questions, New York: Bantam Books, p. 78.

[6] Loeb A. (2021) How Much Time Does Humanity Have Left? *Scientific American*, downloaded from: (https://www.scientificamerican.com/article/how-much-time-does-humanity-have-left/), accessed May 13, 2021.

[7] Schmidt S., Zubrin R. {editors} (1996) Islands in the Sky: Bold New Ideas for Colonizing Space. Wiley.

[8] Cooper J. (2013) Bioterrorism and the Fermi Paradox. *International Journal of Astrobiology* 12(2), pp. 144-148.

[9] Shekhtman L. (2019) NASA Takes a Cue From Silicon Valley to Hatch Artificial Intelligence Technologies. NASA's Goddard Space Flight Center, Greenbelt, MD. Downloaded from: (https://www.nasa.gov/feature/goddard/2019/nasa-takes-a-cue-from-silicon-valley-to-hatch-artificial-intelligence-technologies), accessed May 24, 2021.

[10] Zimmer C. (2011) 100 Trillion Connections: New Efforts Probe and Map the Brain's Detailed Architecture. Scientific American. (https://www.scientificamerican.com/article/100-trillion-connections), accessed May 13, 2021.

[11] Whitwam R. (2013) New Transistor Mimics Human Synapse to Simulate Learning. ExtremeTech.com/Ziff Davis, LLC. (https://www.extremetech.com/extreme/170411-new-transistor-mimics-human-synapse-to-simulate-learning).

[12] National Renewable Energy Laboratory (2021) News Release: Scientists at NREL Report New Synapse-Like Phototransistor. Downloaded from (https://www.nrel.gov/news/press/2021/scientists-at-nrel-report-new-synapse-like-phototransistor.html), accessed April 28, 2021.

[13] Benford J., Benford G. (2013) Starship Century. 2011 100 Year Starship Symposium, Microwave Sciences/Lucky Bat Books.

[14] Bussard R.W. (1960) Galactic Matter and Interstellar Flight. *Astronautica Acta* 6, pp. 179-195.

[15] Whitmire D.P. (1975) Relativistic Spaceflight and the Catalytic Nuclear Ramjet. *Acta Astronautica* 2 (5-6), pp. 497-509.

[16] Moore G.E. (1975) Progress in Digital Integrated Electronics. Technical Digest 1975. *International Electron Devices Meeting*, IEEE, pp. 11-13.

[17] Roser M., Ritchie H. (2013), downloaded from: (https://www.OurWorldInData.org), accessed April 10, 2021.

[18] Waldrop M. (2016) The Chips are Down for Moore's Law. *Nature* 530 (7589), pp. 144-147.

[19] Ahmed I., Aubourg L. (2021) America has Sent Five Rovers to Mars – When Will Humans Follow? Phys.org. Downloaded from: (https://phys.org/news/2021-02-america-rovers-marswhen-humans.html), accessed May 17, 2021.

[20] Ayres R. (1988) Barriers and Breakthroughs: An "Expanding Frontiers" Model of the Technology-Industry Life Cycle. *Technovation* 7(2), pp. 87-115.

[21] NASA (2019) NASA: Moon to Mars. Nasa.gov, Downloaded from (https://www.nasa.gov/specials/moontomars/index.html), accessed May 22, 2021.

[22] Anglada-Escudé G., Amado P.J., Barnes J., et al (2016) A terrestrial planet candidate in a temperate orbit around Proxima Centauri. *Nature* 536 (7617), pp. 437-440.

[23] Gillon M., Triaud A H M J, Demory B. O-, et al (2017) Seven temperate terrestrial planets around the nearby ultracool dwarf star TRAPPIST-1. *Nature* 542 (7642), pp. 456-460.

[24] Zubrin R. (2007) Economic Viability of Mars Colonization, Lockheed Martin Astronautics. Downloaded from: (https://web.archive.org/web/20070928081643/http://www.4frontierscorp.com/dev/assets/Economic%20Viability%20of%20Mars%20Conolozation.pdf), accessed May 23, 2021.

[25] Troutman P.A., Bethke K., Stillwagon F., Caldwell Jr. D.L., Manvi R., Strickland C., Krizan S.A (2003) Revolutionary Concepts for Human Outer Planet Exploration (HOPE). Space Technology & Applications International Forum (STAIF – 2003), "Expanding the Frontiers of Space", downloaded from: (https://ntrs.nasa.gov/api/citations/20030063128/downloads/20030063128.pdf?attachment=true), accessed May 25, 2021.

[26] Brown D., Platt J., Bell, B. (2014) NASA Space Assets Detect Ocean inside Saturn Moon, NASA/JPL, Release 14-099 (April 3, 2014, updated August 7, 2017), Downloaded from: (https://www.nasa.gov/press/2014/april/nasa-space-assets-detect-ocean-inside-saturn-moon), accessed May 25, 2021.

[27] Zubrin R. (1999) Entering Space: Creating a Spacefaring Civilization, *Putnam.*